\documentclass[conference]{IEEEtran}
\IEEEoverridecommandlockouts

\usepackage{cite}
\usepackage{amsmath,amssymb,amsfonts}
\usepackage{algorithmic}
\usepackage{graphicx}
\usepackage{textcomp}
\usepackage{xcolor}
\usepackage[framemethod=TikZ]{mdframed}
\usepackage{xcolor}
\usepackage{marginnote}
\usepackage{pifont}
\usepackage[utf8]{inputenc}
\usepackage{microtype}
\usepackage{inconsolata}
\usepackage{dblfloatfix}
\usepackage{siunitx} 
\usepackage{booktabs} 
\usepackage{tabularx} 
\usepackage{multirow} 
\usepackage{verbatim} 
\usepackage{listings} 
\usepackage{multirow}
\usepackage{booktabs}
\usepackage{subcaption}
\newmdenv[
    leftmargin=10pt,
    rightmargin=10pt,
    backgroundcolor=gray!20,
    linecolor=black,
    linewidth=2pt,
    topline=false,
    bottomline=false,
    skipabove=\baselineskip,
    skipbelow=\baselineskip
]{quotebox}

\usepackage{float} 
\usepackage{multirow}
\usepackage{colortbl}
\usepackage{multirow, colortbl, makecell}

\usepackage{url}  
\usepackage{hyperref}  
\usepackage[table]{xcolor}  

\def\BibTeX{{\rm B\kern-.05em{\sc i\kern-.025em b}\kern-.08em
    T\kern-.1667em\lower.7ex\hbox{E}\kern-.125emX}}
\begin{document}

\title{Quality Over Quantity? LLM-Based Curation \\for a Data-Efficient Audio–Video Foundation Model}

\author{
    \IEEEauthorblockN{Ali Vosoughi$^{*}$\thanks{$^*$ Work completed during an internship at Microsoft Research, Redmond, WA, USA.}}
    \IEEEauthorblockA{\textit{University of Rochester}\\
    Rochester, NY, USA}
\and
\IEEEauthorblockN{Dimitra Emmanouilidou}
\IEEEauthorblockA{\textit{Microsoft Research}\\
Redmond, WA, USA}
\and
\IEEEauthorblockN{Hannes Gamper}
\IEEEauthorblockA{\textit{Microsoft Research}\\
Redmond, WA, USA }
}

\maketitle

\begin{abstract} Integrating audio and visual data for training multimodal foundational models remains a challenge. The Audio-Video Vector Alignment (AVVA) framework addresses this by considering AV scene alignment beyond mere temporal synchronization, and leveraging Large Language Models (LLMs) for data curation.
AVVA implements a scoring mechanism for selecting aligned training data segments. It integrates Whisper, a speech-based foundation model, for audio and DINOv2 for video analysis in a dual-encoder structure with contrastive learning on AV pairs. Evaluations on AudioCaps, VALOR, and VGGSound demonstrate the effectiveness of the proposed model architecture and data curation approach. AVVA achieves a significant improvement in top-k accuracies for video-to-audio retrieval on all datasets compared to DenseAV, while using only 192 hrs of curated training data. 
Furthermore, an ablation study indicates that the data curation process effectively trades data quality for data quantity, yielding increases in top-k retrieval accuracies on AudioCaps, VALOR, and VGGSound, compared to training on the full spectrum of uncurated data. 
\end{abstract}


\section{Introduction and Motivation} \label{sec:intro}

Humans naturally process audiovisual information without any need for textual mediation. For instance, when watching a video, we instinctively merge visual cues with corresponding sounds to create a cohesive understanding of the scene. However, most current multimodal AI systems, like CLIP \cite{radford2021learning} and CLAP \cite{elizalde2023clap}, and majority of other  models~\cite{li2023blip,chen2020uniter, wu2022wav2clip,guzhov2022audioclip,girdhar2023imagebind,vosoughi2024learning,chen2023dialogmcf}, depend on textual captions to connect visual and auditory features. This reliance on text-based alignment is at odds with how humans integrate sensory information, where no explicit textual representation is required.

\begin{figure} \centering \includegraphics[width=\linewidth]{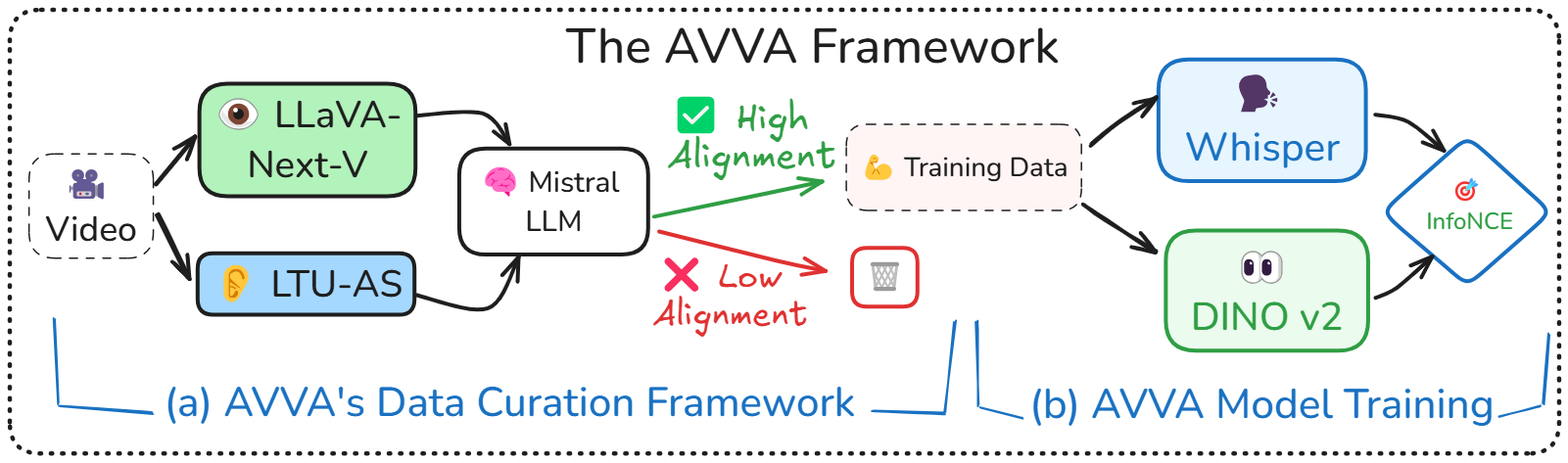} 
    \caption{Overview of the proposed audiovisual alignment approach. (a) Our method's data curation stage uses multimodal reasoning to retain only highly aligned data. It uses  LLaVA-Next-Video \cite{liu2024llavanext, zhang2024llavanextvideo} for video reasoning, LTU-AS \cite{gong2023joint-ltuas} for audio processing, and Mistral \cite{mistral2023} for alignment scoring. (b) AVVA employs Whisper (audio), DINOv2 (video backbone), without the need for textual mediation during  training.}
\label{fig:teaser} 
\end{figure}
Replicating this human-like processing in AI is challenging~\cite{harwath2016unsupervised, tian2020unified,harwath2018jointly}. Existing multimodal models primarily handle individual modalities separately, later merging them based on text-based associations \cite{tang2024video, bi2024eagle, nguyen2024oscar,su2024from,tangsalmonn2024iclr-salmon, kong2024audio,deshmukh2023pengi,gong2024listen,huang2024vtimellm,zhu2024languagebind}. This approach, evident in models like Wav2CLIP \cite{wu2022wav2clip}, AudioCLIP \cite{guzhov2022audioclip} and ImageBind \cite{girdhar2023imagebind}, misses the opportunity to exploit the natural synchronization between audio and visual data. While efforts like AV-HuBERT \cite{shi2022learning-avhubert} and DenseAV \cite{hamilton2024separating} aim to capture linguistic information along with the location of sounds from raw audiovisual pairs, they still rely on speech-image pairs for training, which may restrict their generalization.

To address this gap, we introduce \textbf{AVVA: Audio-Video Vector Alignment}, a framework designed to directly align AV information without any text dependency. The proposed model leverages \textit{Whisper} \cite{radford2023whisper} for audio processing and \textit{DINOv2} \cite{oquab2023dinov2} for visual understanding. This makes AVVA particularly effective in applications requiring concurrent, text-free audiovisual comprehension, such as video analysis and human-computer interaction.  An important data curation stage takes place first, which itself relies on a text-, audio- and video-LLMs.

Our contributions are threefold: (1) AVVA is the first audiovisual foundation model that incorporates a speech foundation model to enable generalized audiovisual representation learning. (2) Unlike previous approaches that often align audio and visual features independently, AVVA introduces a mechanism for joint multimodal reasoning. (3) Our novel data curation mechanism significantly reduces the amount of required training data while still achieving competitive results with state-of-the-art models, which further demonstrates the efficiency and effectiveness of using curated data over the original datasets.

The remainder of the paper is organized as follows:  Section \ref{sec:method} explains our methodology, including data and model design. Section \ref{sec:exps} presents the experimental results, and Section \ref{sec:conclusion} concludes the paper.

\section{AVVA: Audio-Video Vector Alignment} \label{sec:method}
A key feature of the proposed method is the curation and selection of high-quality paired data. AVVA leverages the synergy of three large models—two for multimodal inputs (audio and video) and one for joint reasoning—to compute five alignment scores. These scores will then be used to evaluate the coherence of the audiovisual data. We will explain different parts of the method. More details on the implementation and reproducibility of AVVA, including prompts, statistics of datasets, and ablations studies will be provided at \url{https://github.com/AVVA-curation}.

\subsection{Multimodal Reasoning Engine (MRE)} The potential of most AI techniques for LLMs and multimodal learning often hinges on the diversity and quality of the data they interact with~\cite{sorscher2022beyond,gunasekar2023Phitextbooks}. In this work, we curate the training data via the introduction of a Multimodal Reasoning Engine (MRE), which is a set of prompts for obtaining detailed reasoning of audio, video, and finally to score the level of alignment between audio and video from their textual descriptions,  given a set of five metrics. 

We used multiple audiovisual datasets that cover diverse scenes from both egocentric and exocentric perspectives, and various forms of audio, including natural, music, ambient, and speech, and other complex scenarios. The datasets are:
 Epic-Kitchens~(1.37 hrs)~\cite{damen2020epic}, HowTo100M~(7.77 hrs)~\cite{miech2019howto100m}, Music-MIT~(2.14 hrs)~\cite{zhao2019sound}, VALOR~(train/test 94.57/13.58 hrs)~\cite{chen2023valor}, VGGSound~(train/test 30.23/2.82 hrs)~\cite{chen2020vggsound}, AVE~(10.00 hrs)~\cite{tian2018ave}, AudioSet~(54.83 hrs)~\cite{audioset}, AudioCaps~(train/test 32.64/1.00 hrs)~\cite{kim2019audiocaps}, HD-VILA-100M~(51.45 hrs)~\cite{xue2022hdvila}. All input videos were segmented into 3-sec clips; for longer videos, up to 20 random clips were kept. To achieve joint audio-speech reasoning, each segment was processed using LLaVa-NeXT-Video with LLaMA 3 \cite{liu2024llavanext,zhang2024llavanextvideo} for video reasoning and LTU-AS \cite{gong2023joint-ltuas} with LLaMA 2 for audio reasoning. We used Mistral 7B Instruct v0.3 for prompting and measuring alignment. 

\begin{figure} \centering \includegraphics[width=\linewidth]{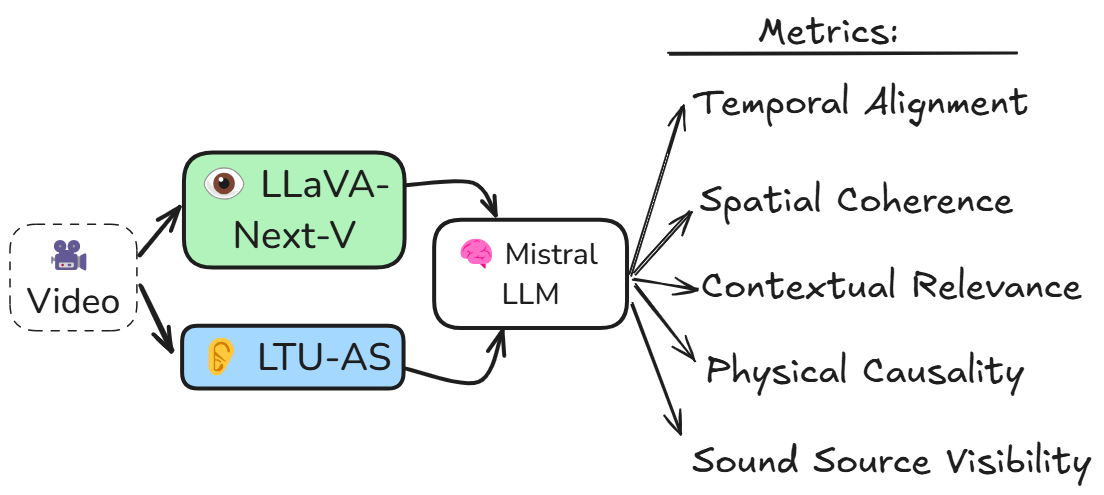}
\caption{The architecture of the MRE. The design integrates outputs of an audio-LLM and a video-LLM into a Mistral LLM to reason over audiovisual scene alignment by integrating 5 alignment scores that were calculated on the AV pairs.} \label{fig:MRE} \end{figure}

Given a video sample, we obtain one caption from LTU-AS describing the audio, and one caption from LLaVA-Next-V describing the video, see Fig. \ref{fig:MRE}. The two captions are then fed to Mistral, along with a prompt request to obtain five separate scores in a scale of [0,10] that aim to capture caption alignment. These scores come from the five metrics: Temporal Alignment, Spatial Coherence, Contextual Relevance, Physical Causality, and Sound Source Visibility. For Temporal Alignment, we ask the system to assess how well the events described in the audio caption match the timing of events in the video caption (e.g., a clap sound syncing with hands meeting). Spatial Coherence evaluates how well the audio description aligns with the spatial layout and objects described in the video (e.g., a car's engine sound moving from left to right as it passes). Contextual Relevance refers to how closely the subject matter and theme of the audio align with those of the video (e.g., kitchen sounds matching cooking activities). Physical Causality assesses the extent to which the described sounds can be logically attributed to the objects, actions, or events depicted in the video (e.g., glass breaking sound matching the visual shattering). Sound Source Visibility considers that some visual objects may produce sound without being visible and others may be visible but silent. 
The prompt details can be found in the Appendix (see GitHub). These alignment scores are then averaged with equal weights, and a final alignment score is produced, with the assumption that a higher score represents better audiovisual alignment. A scoring threshold is subsequently applied for final data curation. 
For reference, retaining  90\% or 70\% of the original training data corresponded to a score threshold of 6.2 and 7.6 respectively.

\subsection{Model Architecture and Language-free Training} 
\begin{figure} \centering \includegraphics[width=\linewidth]{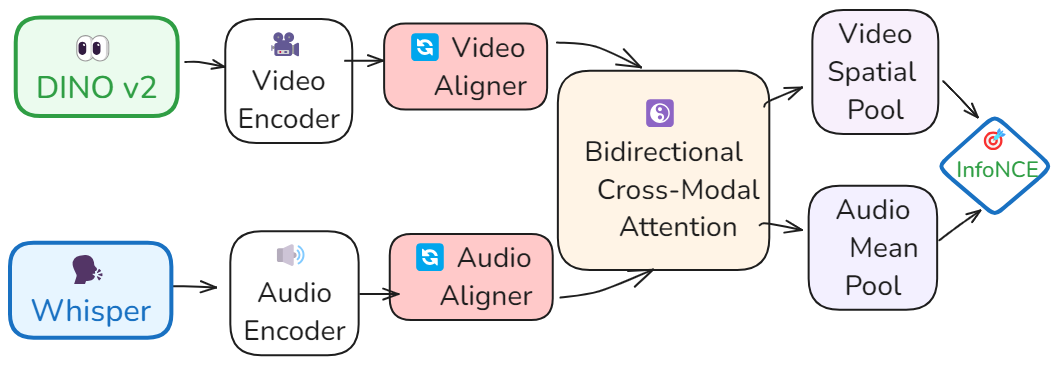} \caption{The AVVA model training. Audio (Whisper) and video (DINOv2) encoders process raw inputs, which are aligned via learnable parameters in aligner layers. A Bidirectional Cross-Modal Attention helps capture the interaction between audio and video features, which are pooled to generate final embeddings for contrastive learning.} \label{fig:net-arch-partial} \end{figure}

The AVVA model employs a bidirectional cross-modal attention mechanism to integrate audio and video modalities using dual encoders—Whisper for audio and DINOv2 for video. We selected DINOv2~\cite{oquab2023dinov2} over models like CLIP~\cite{radford2021learning} due to its ability to capture local visual features, which are crucial for producing high-quality global representations through feature pooling \cite{hamilton2024separating}. For the audio encoder, we utilize Whisper, concatenating 32 layers while discarding the first layer, as applied in \cite{gong2023joint-ltuas}. The architecture is illustrated in Fig. \ref{fig:net-arch-partial}.

The bidirectional attention mechanism, implemented with 8 attention heads and a 768-dimensional hidden state, ensures a robust flow of information between the audio and video streams, treating both modalities as equally important. This design makes AVVA particularly effective for complex multimodal tasks requiring detailed audiovisual understanding, such as synchronized multimedia content generation~\cite{xing2024seeing}, event detection~\cite{berghi2024fusion}, and other tasks requiring detailed audiovisual analysis~\cite{su2023separating}. By aligning the features through learnable aligner layers - implemented as MLPs with dimensions (input\_dim, 1024, 768) and layer normalization, $ReLU$, and dropout 0.2 between layers - and pooling the outputs, the model generates compact embeddings suitable for contrastive learning.

We use the InfoNCE loss function \cite{radford2021learning} with a temperature setting of 0.07 to help the model learn  correlations between audio and video features.
For optimization, we adopt the AdamW optimizer with a learning rate and weight decay set to $10^{-4}$. To maintain computational efficiency and prevent catastrophic forgetting, both the DINOv2 and Whisper encoders are frozen \cite{wang2023makes} during training, focusing on training only the alignment and cross-modal layers.

\section{Experiments}\label{sec:exps}
We conducted three sets of experiments to evaluate the performance of our method in diverse scenarios.

\subsection{Cross-modal Retrieval}
In this experiment, we assess the ability of AVVA to retrieve audio from video input and vice versa, across three datasets: AudioCaps~\cite{kim2019audiocaps}, VALOR~\cite{chen2023valor}, and VGG-Sound~\cite{chen2020vggsound}. Each test is performed on 3-second video segments containing embedded audio, and we compare our model against Wav2CLIP~\cite{wu2022wav2clip}, DenseAV~\cite{hamilton2024separating}, Random weights, and ImageBind~\cite{girdhar2023imagebind}. Each retrieval test is evaluated K=100 times on N=100 random audio/video files per iteration. No duplication occurs within each set of 100 samples per run. Results are reported as statistical averages.

AVVA achieves audio-to-video accuracy comparable to DenseAV, with significantly improved video-to-audio accuracy, despite using only 192 hrs of carefully curated data compared to DenseAV's 5,800 hrs, demonstrating a 30x improvement in data efficiency. This showcases the effectiveness of high-quality, curated audiovisual pairs curated by our system.
Notably, all competing methods in Table~\ref{tab:main-paper-performance} were trained on larger datasets. For instance, Wav2CLIP was trained on approximately 278 hrs of data. This comparison highlights the impact of effective data curation on enhancing model performance. The results for AVVA in Table~\ref{tab:main-paper-performance} reflect an MRE threshold score of 7.6 out of 10, based on selecting the epoch with the minimum validation los.
A key observation from our experiments is that increasing the amount of training data does not always lead to better performance. While more data should generally improve model accuracy initially, adding data can introduce noise, particularly when the additional data is less curated or includes irrelevant or misaligned audiovisual pairs. This phenomenon is portrayed in our experiments, where models trained on large but uncurated datasets such as Wav2CLIP and DenseAV performed equivalent or worse than AVVA, especially in V2A retrieval, despite having access to more data.

\begin{figure*}[t]
    \centering
    \begin{subfigure}[b]{0.85\textwidth}
        \includegraphics[width=\linewidth]{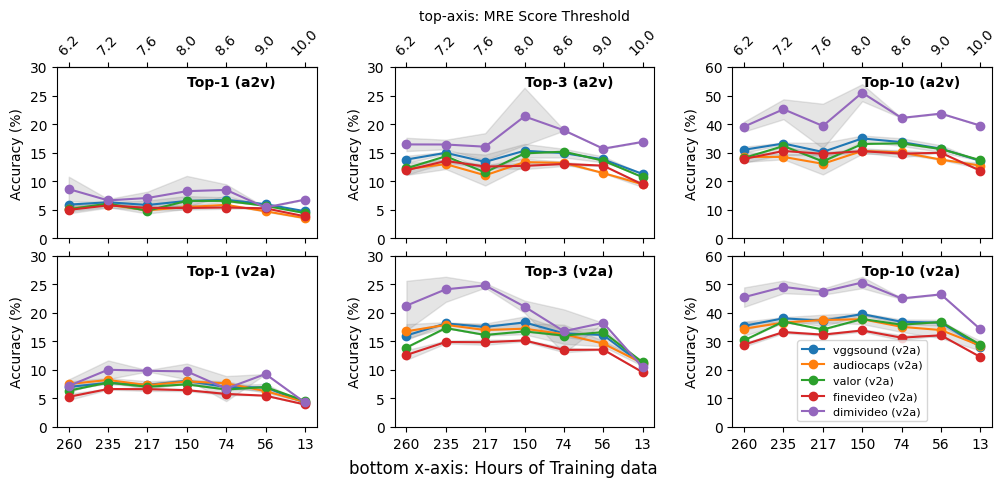}
        \label{fig:audiocaps_performance_top1}
    \end{subfigure}
    \caption{Audio-to-video model performance over hours of training data, as determined by varying the selection of the MRE score threshold, shown for Top-$k=\{1,3,10\}$ accuracies. 
    }
    \label{fig:audiocaps_performance}
\end{figure*}

\begin{table*}
\centering
\caption{\textbf{Performance on Audio (A) - Video (V) Retrieval} (Top-$k=\{1,3,10\}$) ~(\%). Standard deviations shown as superscripts depict performance variation over K=100 retrieval repetitions of random test subsets of size N=100. }
\label{tab:main-paper-performance}
\renewcommand{\arraystretch}{1.37}  
\scalebox{0.84}{
\begin{tabular}{|l|c|ccc|ccc|ccc|}
\hline
\multirow{2}{*}{\textbf{Method}} & \multirow{2}{*}{\textbf{Retrieval Type}} 
& \multicolumn{3}{c|}{\textbf{AudioCaps}} 
& \multicolumn{3}{c|}{\textbf{VALOR}} 
& \multicolumn{3}{c|}{\textbf{VGG-Sound}} \\
 &  & \textbf{Top1} & \textbf{Top3} & \textbf{Top10} & \textbf{Top1} & \textbf{Top3} & \textbf{Top10} & \textbf{Top1} & \textbf{Top3} & \textbf{Top10} \\
\hline
\multirow{2}{*}{Wav2CLIP\cite{wu2022wav2clip}} & A\(\to\)V 
& $1.20^{\pm 0.98}$ & $6.40^{\pm 3.01}$ & $18.60^{\pm 3.83}$ 
& $3.60^{\pm 0.80}$ & $8.60^{\pm 0.49}$ & $18.20^{\pm 3.54}$ 
& $3.40^{\pm 1.36}$ & $8.20^{\pm 1.17}$ & $19.80^{\pm 3.06}$ \\
 & V\(\to\)A 
& $3.80^{\pm 2.14}$ & $10.00^{\pm 3.22}$ & $20.00^{\pm 3.63}$ 
& $4.20^{\pm 2.64}$ & $8.00^{\pm 4.24}$ & $19.00^{\pm 3.52}$ 
& $3.80^{\pm 1.94}$ & $9.20^{\pm 2.32}$ & $19.60^{\pm 1.62}$ \\
\hline
\multirow{2}{*}{Random} & A\(\to\)V 
& $1.40^{\pm 0.49}$ & $3.80^{\pm 0.75}$ & $11.80^{\pm 1.17}$
& $1.20^{\pm 0.75}$ & $3.20^{\pm 0.40}$ & $11.60^{\pm 1.62}$ 
& $1.20^{\pm 0.40}$ & $3.40^{\pm 0.49}$ & $11.60^{\pm 2.15}$ \\
 & V\(\to\)A 
& $1.00^{\pm 0.00}$ & $3.60^{\pm 0.80}$ & $10.80^{\pm 1.17}$ 
& $1.20^{\pm 0.40}$ & $3.20^{\pm 0.75}$ & $11.00^{\pm 0.63}$ 
& $1.00^{\pm 0.00}$ & $3.00^{\pm 0.00}$ & $10.60^{\pm 0.80}$ \\
\hline
\multirow{2}{*}{DenseAV\cite{hamilton2024separating}} & A\(\to\)V 
& $10.20^{\pm 2.04}$ & $22.60^{\pm 4.54}$ & $49.40^{\pm 4.54}$ 
& $7.80^{\pm 5.19}$ & $19.00^{\pm 5.90}$ & $41.80^{\pm 4.79}$ 
& $6.80^{\pm 2.64}$ & $16.00^{\pm 2.90}$ & $43.20^{\pm 3.43}$ \\
 & V\(\to\)A 
& $1.40^{\pm 0.80}$ & $5.60^{\pm 1.85}$ & $26.40^{\pm 2.73}$ 
& $2.20^{\pm 1.17}$ & $5.80^{\pm 2.79}$ & $24.60^{\pm 7.68}$ 
& $1.60^{\pm 1.02}$ & $5.00^{\pm 0.63}$ & $22.60^{\pm 2.58}$ \\
\hline
\multirow{2}{*}{ImageBind\cite{girdhar2023imagebind}} & A\(\to\)V 
& ${62.00^{\pm 2.28}}$ & ${83.40^{\pm 3.01}}$ & ${92.60^{\pm 1.85}}$ 
& ${55.80^{\pm 4.66}}$ & ${71.60^{\pm 3.61}}$ & ${85.00^{\pm 3.74}}$ 
& ${50.60^{\pm 3.14}}$ & ${74.00^{\pm 5.93}}$ & ${88.20^{\pm 2.99}}$ \\
 & V\(\to\)A 
& ${64.00^{\pm 5.37}}$ & ${85.40^{\pm 4.27}}$ & ${95.40^{\pm 0.80}}$ 
& ${58.80^{\pm 4.71}}$ & ${73.60^{\pm 4.36}}$ & ${86.60^{\pm 3.20}}$ 
& ${53.20^{\pm 3.31}}$ & ${73.40^{\pm 6.02}}$ & ${85.60^{\pm 3.20}}$ \\
\hline
\multirow{2}{*}{\textbf{AVVA (Ours)}} & A\(\to\)V 
& ${6.57^{\pm 2.30}}$ & ${13.84^{\pm 2.80}}$ & ${31.68^{\pm 3.57}}$ 
& ${6.69^{\pm 2.13}}$ & ${15.63^{\pm 3.52}}$ & ${33.67^{\pm 4.40}}$ 
& ${6.71^{\pm 1.91}}$ & ${15.02^{\pm 2.73}}$ & ${33.86^{\pm 4.23}}$ \\ 
 & V\(\to\)A
& ${6.23^{\pm 2.09}}$ & ${14.70^{\pm 3.17}}$ & ${31.06^{\pm 3.52}}$ 
& ${7.75^{\pm 2.61}}$ & ${16.64^{\pm 3.65}}$ & ${34.27^{\pm 4.71}}$ 
& ${6.86^{\pm 2.34}}$ & ${14.47^{\pm 3.15}}$ & ${32.84^{\pm 3.89}}$ \\
\hline
\end{tabular}
}
\end{table*}

\subsection{Data Curation Impact on Performance}
This experiment evaluates the effect of data curation on model performance in cross-modal retrieval tasks. As illustrated in Fig. \ref{fig:audiocaps_performance}, higher curation thresholds lead to improved performance. We argue that meaningful data curation reduces noise in training data, allowing the model to focus on high-quality examples, resulting in more accurate retrieval across modalities. Similar plots were obtained for the other test sets and the video-to-audio retrieval task. Despite being computationally expensive - typically increasing preprocessing time by 6 seconds per GPU time per segment - the improved data quality curation enhances the model's ability to generalize, which showcases the importance of the data selection in multimodal training. The findings are summarized in Table \ref{tab:performance_increase}, in terms of performance improvement (\%) as compared to training on full data . AVVA achieves top-1 performance increases relative to original dataset with same hrs of training across datasets for both audio-to-video and video-to-audio tasks. For audio-to-video retrieval, AVVA achieves increases of 18.0, 16.24, and 13.57 percentage points (\%) in top 1, 3, and 10 for AudioCaps; for VALOR, increases of 22.67, 23.97, and 15.42 \% respectively; and for VGGSound, increases of 23.25, 15.79, and 10.44 \% in top 1, 3, and 10. The proposed method also shows merit in the V2A task, in this case more moderate improvements than for the A2V task,  shown in the second column of the Table.

\begin{table}[H]
\centering
\caption{Performance increases in cross-modal retrieval tasks with data curation. Top-$k=\{1,3,10\}$ \% increase across various datasets as compared to training on full original data.}
\label{tab:performance_increase}
\scalebox{0.95}{
\begin{tabular}{|l|ccc|ccc|}
\hline
\multirow{2}{*}{\textbf{Dataset}} & \multicolumn{3}{c|}{\cellcolor[gray]{0.9}\textbf{Audio-to-Video ↑ (\%)}} & \multicolumn{3}{c|}{\cellcolor[gray]{0.9}\textbf{Video-to-Audio ↑ (\%)}} \\
& \cellcolor[gray]{0.9}\textbf{Top 1} & \cellcolor[gray]{0.9}\textbf{Top 3} & \cellcolor[gray]{0.9}\textbf{Top 10} & \cellcolor[gray]{0.9}\textbf{Top 1} & \cellcolor[gray]{0.9}\textbf{Top 3} & \cellcolor[gray]{0.9}\textbf{Top 10} \\
\hline
AudioCaps & \cellcolor[gray]{0.95}18.0 & 16.24 & \cellcolor[gray]{0.95}13.57 & 11.08 & \cellcolor[gray]{0.95}11.29 & 14.71 \\
VALOR & 22.67 & \cellcolor[gray]{0.95}23.97 & 15.42 & \cellcolor[gray]{0.95}10.44 & 4.00 & \cellcolor[gray]{0.95}8.50 \\
VGGSound & \cellcolor[gray]{0.95}23.25 & 15.79 & \cellcolor[gray]{0.95}10.44 & 1.76 & \cellcolor[gray]{0.95}3.41 & 1.86 \\
\hline
\end{tabular}
}
\end{table}

\subsection{Temporal Alignment }
\begin{figure}
    \centering
    \includegraphics[width=1.05\linewidth]{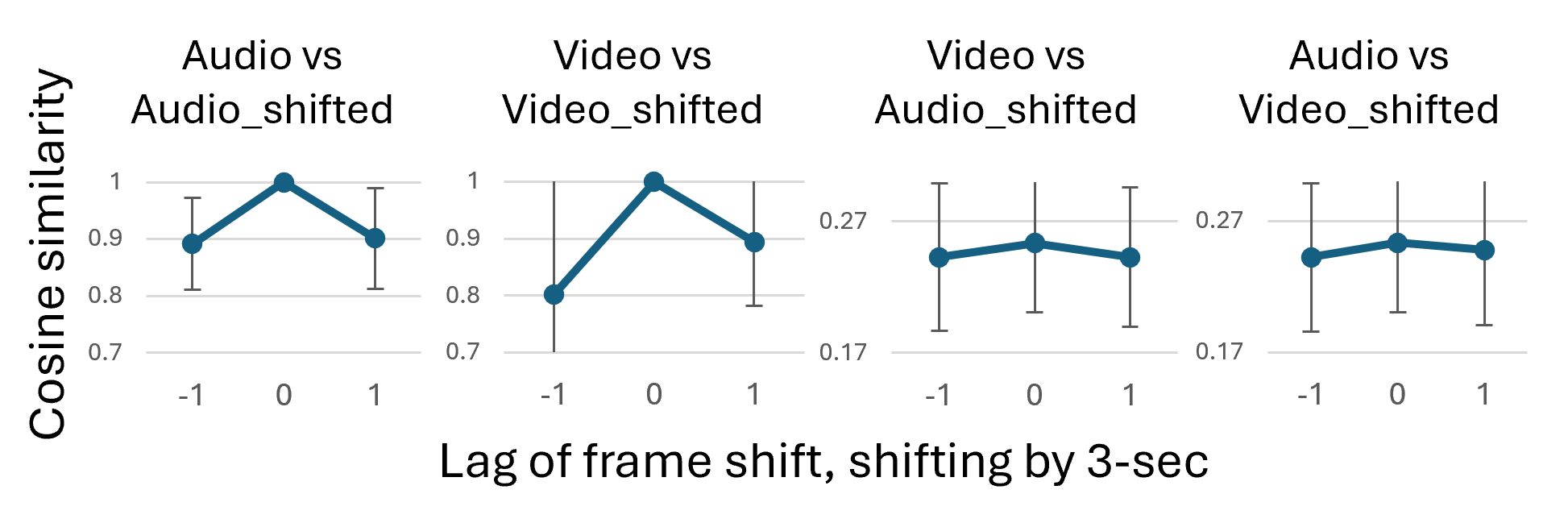}
    \caption{Cosine similarity between AVVA embeddings as a function of audio shifts and video shifts. The data points show mean similarity scores at each shift level, comparing audio embeddings against shifted audio (panel 1) and shifted video (panel 4); and vice versa, similarity of video embeddings against shifted video (panel 2) and shifted audio (panel 3). 
    }
    \label{fig:similarity_shifts}
\end{figure}

To investigate the audio-video temporal alignment, we conducted simulations where audio segments were systematically shifted relative to their corresponding video segments across 3-sec increments. For each shift, cosine similarity between audio and video embeddings was computed to assess the alignment quality. The multimodal audiovisual embeddings were extracted using our pre-trained model. 
Figure \ref{fig:similarity_shifts} shows average cosine similarity scores for the video and audio embeddings of 200 samples across both audio and video shifts. The analysis reveals a clear peak in similarity at  0 sec shift between audio and video, providing evidence of meaningful audio-video learning.

It is important to consider the nature of the data when interpreting these results. Events like hammering or gunshots, which involve sharp and temporally precise correlations between sound and image, exhibit a different behavior compared to more continuous or slower-changing video scenes. For example, considering a video of a train moving in the distance, subtle audio-video delays may be less perceptually disruptive but still affect cosine similarity scores. In such cases, lower cosine similarity  may not necessarily imply poor alignment but rather may reflect the characteristics of the content, emphasizing the critical role of data context in assessing audiovisual alignment.

\section{Conclusion}
\label{sec:conclusion}
AVVA addresses the challenges of joint multimodal learning by directly processing and curating multi-faceted aligned AV data without linguistic mediation in model training. Our approach, utilizing a speech foundation model backbone, demonstrates significant improvements in AV retrieval tasks.
The LLM-based MRE module for data curation rejects audiovisual pairs of low-scoring alignment and helps the model achieve comparable performance to state-of-the-art methods with substantially less training data. AVVA matches DenseAV's performance using only $\sim$ 192 hrs of curated data, compared to DenseAV's 5800+ hrs – a 30x gain in data utilization.
Experiments across multiple datasets showcase the merit of AVVA's methodology on reducing data utilization while maintaining or improved performance. While more work is needed to render the data curation process less computationally expensive, including more efficient reasoning engines, the five metrics comprising the proposed MRE score show a lot of promise, and the overall results highlight the importance of data quality in advancing multimodal AI models.

\vfill\pagebreak

\bibliographystyle{myIEEE}
\bibliography{refs}


\end{document}